\documentclass[trackchanges,twocolumn]{aastex7}

\usepackage{threeparttable}

\begin{document}

\title{A magnetic white dwarf formed through a binary merger within 35 million years}

\author[]{Huahui Yan}
\affiliation{School of Physics and Astronomy, Sun Yat-sen University, DaXue Road, Zhuhai 519082, GuangDong, People’s Republic of China.}
\affiliation{CSST Science Center for the Guangdong-Hongkong-Macau Greater Bay Area, Sun Yat-sen University, DaXue Road, Zhuhai 519082, GuangDong, People’s Republic of China}
\email[show]{}  

\author[]{Jiamao Lin}
\affiliation{School of Physics and Astronomy, Sun Yat-sen University, DaXue Road, Zhuhai 519082, GuangDong, People’s Republic of China.}
\affiliation{CSST Science Center for the Guangdong-Hongkong-Macau Greater Bay Area, Sun Yat-sen University, DaXue Road, Zhuhai 519082, GuangDong, People’s Republic of China}
\email[show]{}  

\author[]{Rizhong Zheng}
\affiliation{Yunnan Observatories, Chinese Academy of Sciences, KunMing 650216, YunNan, People’s Republic of China.}
\affiliation{University of Chinese Academy of Sciences, Beijing 100049, People’s Republic of China}
\email[show]{} 

\author[]{Li Wang}
\affiliation{School of Physics and Astronomy, Sun Yat-sen University, DaXue Road, Zhuhai 519082, GuangDong, People’s Republic of China.}
\affiliation{CSST Science Center for the Guangdong-Hongkong-Macau Greater Bay Area, Sun Yat-sen University, DaXue Road, Zhuhai 519082, GuangDong, People’s Republic of China}
\email[show]{} 

\author[]{Genghao Liu}
\affiliation{School of Physics and Astronomy, Sun Yat-sen University, DaXue Road, Zhuhai 519082, GuangDong, People’s Republic of China.}
\affiliation{CSST Science Center for the Guangdong-Hongkong-Macau Greater Bay Area, Sun Yat-sen University, DaXue Road, Zhuhai 519082, GuangDong, People’s Republic of China}
\email[show]{}     

\author[]{Liangliang Ren}
\affiliation{School of Electrical and Electronic Engineering, Anhui Science and Technology University, BengBu 233030, AnHui, People’s Republic of China.}
\email[show]{}    

\author[]{Zhen Guo}
\affiliation{Instituto de F{\'i}sica y Astronom{\'i}a, Universidad de Valpara{\'i}so, ave. Gran Breta{\~n}a, 1111, Casilla 5030, Valpara{\'i}so 2520000, Valpara{\'i}so, Chile.}
\affiliation{Millennium Institute of Astrophysics, Nuncio Monse{\~n}or Sotero Sanz 100, Of. 104, Providencia, Santiago, Chile.}
\affiliation{Centre for Astrophysics Research, University of Hertfordshire, College Lane, Hatfield AL10 9AB, England, UK.}
\email[show]{}     

\author[]{Siyi Xu}
\affiliation{Gemini Observatory/NSF’s NOIRLab, 670 N. A’ohoku Place, 96720, Hawaii, United State.}
\email[show]{}  

\author[]{Zhangliang Chen}
\affiliation{School of Physics and Astronomy, Sun Yat-sen University, DaXue Road, Zhuhai 519082, GuangDong, People’s Republic of China.}
\affiliation{CSST Science Center for the Guangdong-Hongkong-Macau Greater Bay Area, Sun Yat-sen University, DaXue Road, Zhuhai 519082, GuangDong, People’s Republic of China}
\email[show]{}  

\author[]{Chun Chen}
\affiliation{School of Physics and Astronomy, Sun Yat-sen University, DaXue Road, Zhuhai 519082, GuangDong, People’s Republic of China.}
\affiliation{CSST Science Center for the Guangdong-Hongkong-Macau Greater Bay Area, Sun Yat-sen University, DaXue Road, Zhuhai 519082, GuangDong, People’s Republic of China}
\email[show]{}  

\author[]{Bo Ma}
\affiliation{School of Physics and Astronomy, Sun Yat-sen University, DaXue Road, Zhuhai 519082, GuangDong, People’s Republic of China.}
\affiliation{CSST Science Center for the Guangdong-Hongkong-Macau Greater Bay Area, Sun Yat-sen University, DaXue Road, Zhuhai 519082, GuangDong, People’s Republic of China}
\email[show]{}

\author[]{Yong Shao}
\affiliation{Department of Astronomy, NanJing University, Xianlin Street, NanJing 210023, JiangSu, People’s Republic of China.}
\email[show]{}  

\author[]{Zhenwei Li}
\affiliation{Yunnan Observatories, Chinese Academy of Sciences, KunMing 650216, YunNan, People’s Republic of China.}
\email[show]{} 

\author[]{Xianfei Zhang}
\affiliation{Institute for Frontiers in Astronomy and Astrophysics, Beijing Normal University, No.19, Xinjiekouwai Street, BeiJing 100875, BeiJing, People’s Republic of China.}
\affiliation{School of Physics and Astronomy, Beijing Normal University, No.19, Xinjiekouwai Street, BeiJing 100875, BeiJing, People’s Republic of China}
\email[show]{}

\author[]{Christoffer Fremling}
\affiliation{Caltech Optical Observatories, California Institute of Technology, Pasadena, CA 91125, USA.}
\affiliation{Division of Physics, Mathematics and Astronomy, California Institute of Technology, Pasadena, CA 91125, USA.}
\email[show]{}

\author[]{Jan J. Eldridge}
\affiliation{Department of Physics, University of Auckland, Private Bag 92019, Auckland, New Zealand}
\email[show]{}

\author[]{Hongwei Ge}
\affiliation{Yunnan Observatories, Chinese Academy of Sciences, KunMing 650216, YunNan, People’s Republic of China.}
\affiliation{Key Laboratory for Structure and Evolution of Celestial Objects, Chinese Academy of Sciences,KunMing 650216, YunNan, People’s Republic of China.}
\affiliation{International Centre of Supernovae, Yunnan Key Laboratory, KunMing 650216, YunNan, People’s Republic of China.}
\email[show]{}  

\correspondingauthor{Chengyuan Li,\,Hongwei Ge}
\author[]{Chengyuan Li}
\affiliation{School of Physics and Astronomy, Sun Yat-sen University, DaXue Road, Zhuhai 519082, GuangDong, People’s Republic of China.}
\affiliation{CSST Science Center for the Guangdong-Hongkong-Macau Greater Bay Area, Sun Yat-sen University, DaXue Road, Zhuhai 519082, GuangDong, People’s Republic of China}
\email[show]{lichengy5@mail.sysu.edu.cn;\,gehw@ynao.ac.cn}








\begin{abstract}
White dwarfs (WDs) represent the final evolutionary stage of most stars, typically originating from progenitor stars with masses below approximately 8 $M_{\odot}$ to 10 $M_{\odot}$. Formation through single-star evolution generally requires at least 25 Myr, with the youngest WDs often near the Chandrasekhar limit of 1.4 $M_{\odot}$. In contrast, WDs formed via binary channels, such as mergers or mass transfer, can develop smaller masses in a shorter timescale and may exhibit unique characteristics, including strong surface magnetic fields and rapid rotation. Accurately determining the ages of these WDs is essential for understanding their formation. A valuable method involves studying WDs in star clusters, where member stars share the same age and chemical composition, allowing for precise constraints on the formation times and metallicities of the WDs’ progenitors. Here we report a WD found in the open cluster RSG 5, which is only 35 Myr old. The WD's mass is lower than 1.05 $M_{\odot}$, indicating it may not have formed through single-star evolution.  The WD possesses an exceptionally strong surface magnetic field ($\ge$ 200 MG), a short rotational period ($\sim$6.5 min), and, most notably, a co-rotating half-ring of ionized circumstellar debris. This distinctive feature provides evidence for a binary merger origin, a scenario further substantiated by our stellar evolution models.

\end{abstract}



\section{Introduction} 
When a star with mass below 8--10 $M_{\odot}$ dies, its carbon-oxygen (sometimes neon) core collapses into a white dwarf (WD). WDs are the most common stellar remnants, comprising 5--7\% of local stars by volume \citep{Gaia1Collaboration2021A&A...649A...6G}. 
WDs are distinguished by their Earth-like radii and low luminosities, with a mass limit of 1.44 $M_{\odot}$ \citep{Koester2013pss4.book..559K}. These massive WDs typically require at least 25 Myr to form \citep{Limongi2024ApJS..270...29L}. If WDs undergo binary interactions, they can produce extreme phenomena like Type Ia SNe, X-ray binaries, and gravitational waves. Such WDs often show unusual properties, including strong magnetic fields, rapid rotation, and accretion emission lines \citep{Tout2008MNRAS.387..897T,Garc2012ApJ...749...25G,Nordhaus2011PNAS..108.3135N}. Additionally, their formation timescales may differ significantly from those of typical WDs \citep{Marsh1995MNRAS.275..828M,Iben1990ApJ...353..215I}. 
While many WDs likely form through binary interactions, no observations strictly constrain their evolutionary tracks (i.e., when and how interactions occurred). Precise formation times are needed, but determining individual stellar ages remains challenging.


If a WD is found to belong to a star cluster, this information is highly valuable. Stars in a cluster usually form at the same time from the same molecular cloud, sharing similar ages and chemical compositions \citep{BruzualA2010RSPTA.368..783B}, which would greatly improve our understanding of the WD's evolutionary path. Under the framework of single-star evolution, if the evolutionary age of a WD substantially exceeds that of its associated star cluster, it indicates that the WD may have undergone binary interactions, such as mass transfer or binary mergers. Thanks to the high-precision astrometry and photometry from Gaia mission \citep{GaiaCollaboration2016A&A...595A...1G}, the number of WDs with accurate parallax measurements has increased from about 200 before Gaia \citep{Bergeron2001ApJS..133..413B,Subasavage2017AJ....154...32S,Bedard2017ApJ...848...11B} to around 350,000 in the Gaia early data release 3 (EDR3) and DR3 \citep{Gaia1Collaboration2021A&A...649A...6G,2021MNRASGentileFusillo}. A large sample of WDs enables us to effectively identify those belonging to star clusters.

Here, we report the discovery of a WD in the open cluster (OC) RSG 5, which is only 35 Myr-old. With a mass of about 1.05 $M_{\odot}$, it could not have formed through the single-star evolution channel within the age of the cluster. It has an exceptionally strong surface magnetic field ($\ge$ 200 MG), a rapid rotation period (only $\sim$6.5 minutes), and a co-rotating, half-ring of ionized circumstellar debris, hinting that it may have formed through binary evolution.

\section{Identification of a cluster white dwarf member}\label{sec:style}

\begin{figure*} 
	\centering
	\includegraphics[width=0.49\textwidth]{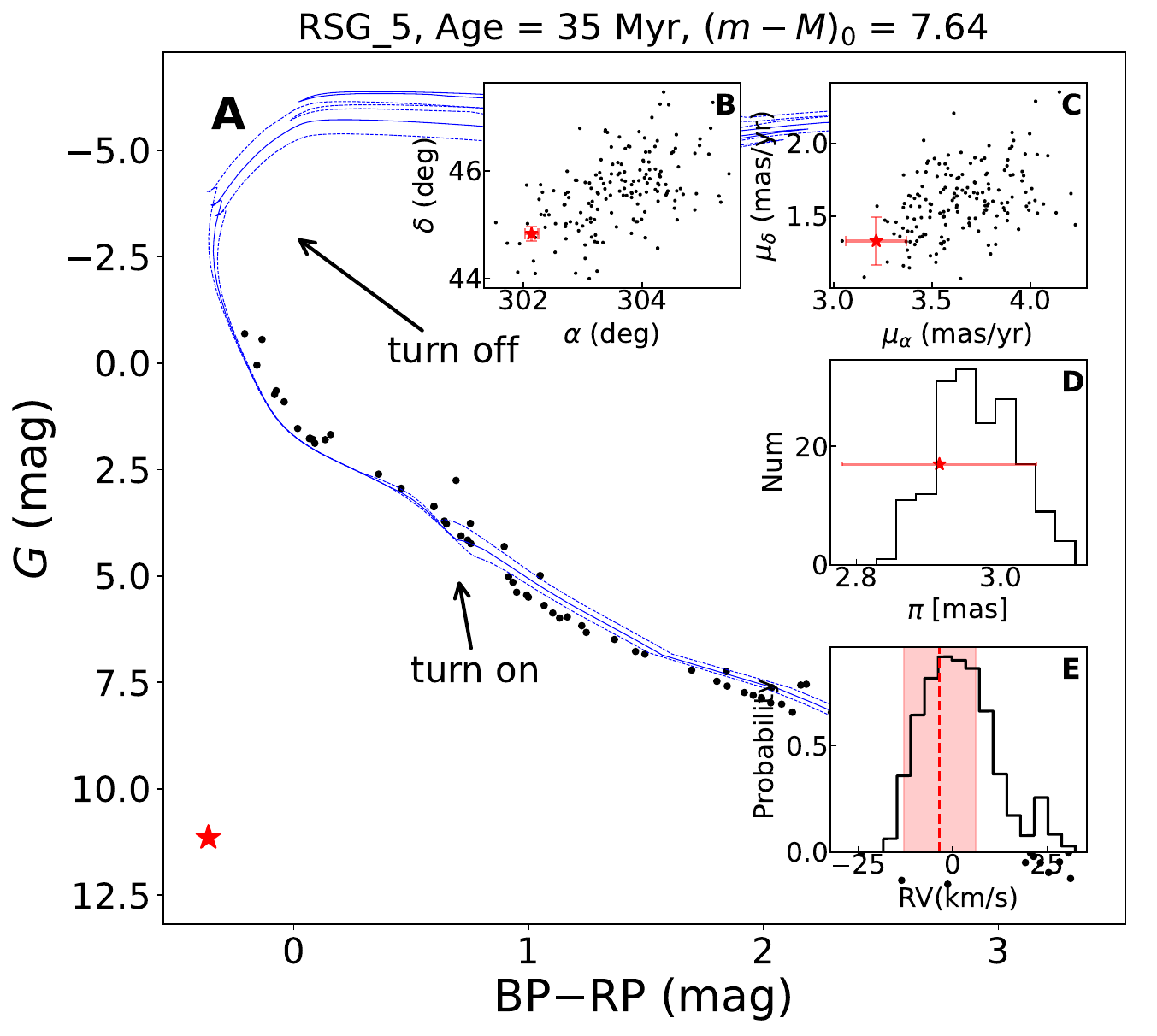} 
    \includegraphics[width=0.49\textwidth]{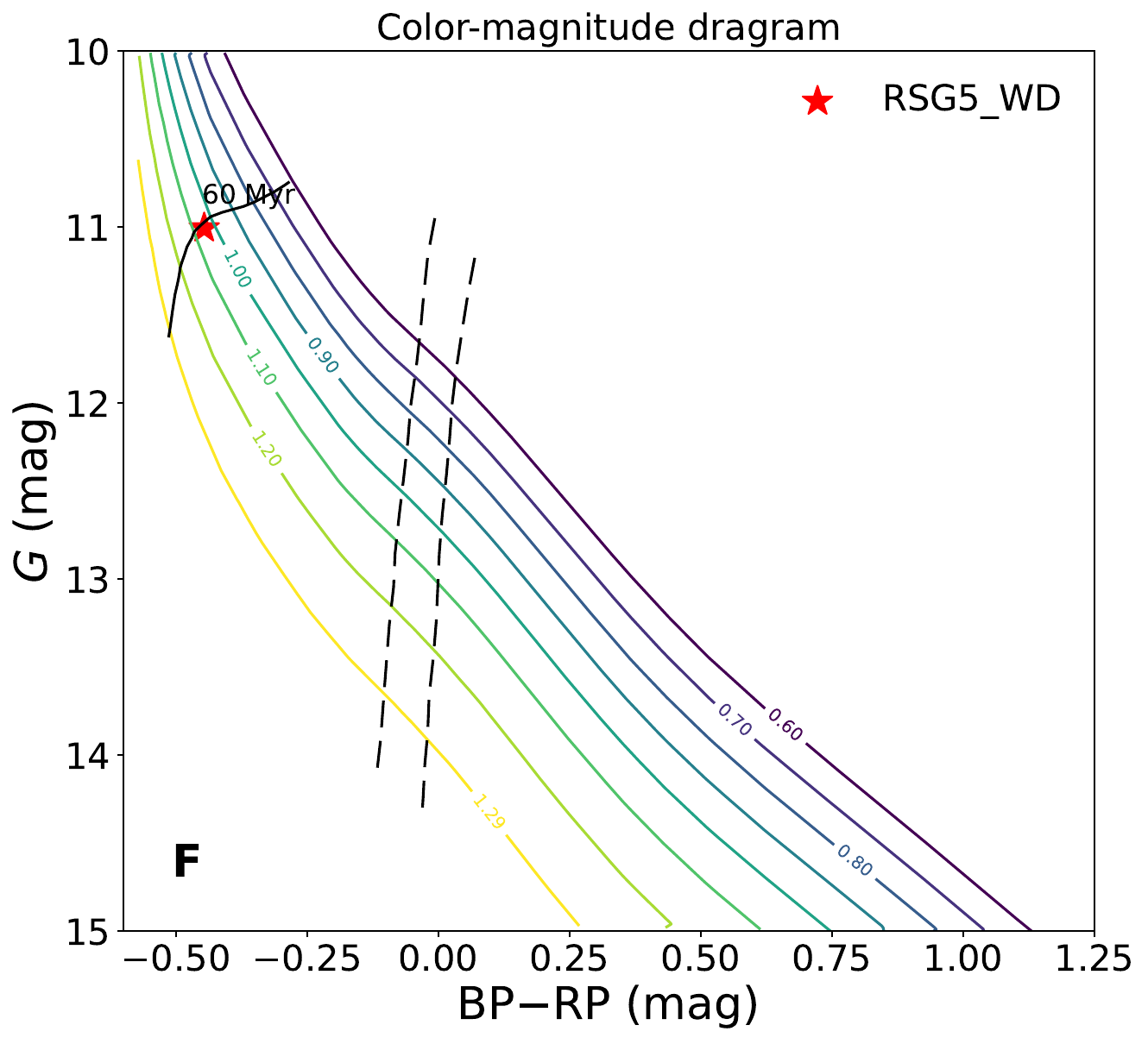}
    \caption{Gaia astrometric parameters and color-magnitude diagram for RSG 5 and RSG5-WD. Left panel: (\textbf{A}): The color-magnitude diagram ($BP-RP$ vs. $G$ band) of the star cluster with isochrone fits. The solid blue line represents the fit for an age of 35 million years (Myr), while the dashed blue lines represent the fit for ages of 30 Myr and 40 Myr. The black dots represent the member stars of the star cluster, while the red pentagram represents RSG5-WD. The black arrows indicate the positions of the main sequence turn-off and the main sequence turn-on. (\textbf{B}): Spatial coordinate distribution of RSG 5 members and RSG5-WD. (\textbf{C}): Proper motion distribution of RSG 5 members and RSG5-WD.  (\textbf{D}): Parallax distribution of RSG 5 members and RSG5-WD. 
    (\textbf{E}): The histogram represents the probability distribution of RV of the magnetically trapped half-ring of ionized circumstellar debris, calculated in \cite{cristea2025halfringionizedcircumstellarmaterial}.
    The red dashed line indicates the average RV of the RSG 5 member stars, and the shaded area represents the 3$\sigma$ level uncertainty.
    Right panel: (\textbf{F}): The solid lines represent the theoretical cooling tracks for WDs with masses ranging from 0.6 M$_{\odot}$ to 1.28 M$_{\odot}$, while the black solid line indicates the 60 Myr cooling age track, both based on a thick-hydrogen carbon-oxygen model. The vertical dashed lines mark the boundaries of the ZZ Ceti instability strip.}

	\label{fig:1} 
\end{figure*}

We conducted a systematic search for WDs in young OCs, utilizing the spatial and kinematic information of OCs provided by \cite{Hunt2024A&A...686A..42H}, combined with the WD database identified by \cite{2021MNRASGentileFusillo} based on Gaia DR3 data. Detailed search methods are provided in Appendix \ref{search method}. A total of 439 WD member candidates associated with 117 OCs are identified. Among them, one (RSG5-WD; Gaia DR3 2082008971824158720) shares equatorial coordinates, parallax, and proper motion with cluster RSG5. It exhibits a magnetically trapped half-ring of ionized circumstellar debris \citep{cristea2025halfringionizedcircumstellarmaterial}.
The radial velocity (RV, $0.8^{+9.6}_{-8.9}$ km s$^{-1}$) of the half-ring of ionized circumstellar debris is also consistent with the cluster's average RV ($-3.4\pm3.2$ km s$^{-1}$; \citep{Hunt2023A&A...673A.114H}). If we consider this WD and the ionized circumstellar debris as a rigid system, it is reasonable to regard RSG5-WD as a cluster member that is spatially and kinematically coeval (see panels B--E of Fig.\ref{fig:1}). Its cluster membership is supported by unsupervised machine learning, as well \citep{Bouma2022AJ....164..215B,Hunt2023A&A...673A.114H,  Prisegen2023A&A...678A..20P}.

\subsection{Age Determination}

RSG 5 is an exceptionally young cluster discovered by \cite{Roser2016A&A...595A..22R}, with age estimates spanning 22-57 Myr \citep{Hunt2024A&A...686A..42H, Almeida2023MNRAS.525.2315A, Bouma2022AJ....164..215B}. We fitted RSG 5's CMD using PARSEC isochrones \citep{Nguyen2022A&A...665A.126N}, with Gaia-band extinction coefficients from $R_V = 3.1$ curves \citep{Cardelli1989ApJ...345..245C, ODonnell1994ApJ...422..158O}. We constrain RSG 5's age primarily using turn-on stars, with their CMD positions marked by black arrows in Fig.~\ref{fig:1} (panel \textbf{A}). We determined the best-fitting PARSEC isochrone by matching the blue edge of the turn-off and turn-on point, selecting from ages 20--100 Myr (1 Myr steps). The optimal fit (35$\pm$5 Myr) has $Z=0.015$ and $(m-M)_{0}=7.64$ mag ($d=337$ pc).

We obtained low-resolution spectra using the Low Resolution Imaging Spectrometer (LRIS) on Keck I \citep{Oke1995PASP..107..375O} (Program IDs C266/C267). Data were reduced with \texttt{PypeIt} \citep{Prochaska2020zndo...3743493P}, a Python-based spectroscopic reduction pipeline. A blackbody fit to RSG5-WD's spectrum (Program ID C267) gives an effective temperature $T_{\rm eff} = 32,\!190^{+4,\!390}_{-3,\!490}$ K (Fig.~\ref{fig:s1}, Appendix). The high temperature ($T_{\rm eff} \approx 32,\!190$ K) excludes a DC WD classification ($T_{\rm eff} \lesssim 10,\!000$ K; \citep{Kilic2025ApJ...979..157K}). We detect Zeeman-split Balmer lines, suggesting RSG5-WD as a magnetic DA WD. 

We employed theoretical cooling tracks from \cite{Bedard2020ApJ...901...93B}, correcting $BP-RP$ and $G$-band extinction using \cite{2021MNRASGentileFusillo}. Then, we used Markov Chain Monte Carlo (MCMC) method to compute the $T_{\mathrm{eff}}$, $\log{\mathrm{g}}$ and mass of RSG5-WD. These results (mass, $T_{\mathrm{eff}}$, $\log{\mathrm{g}}$ ) are presented in Fig.~\ref{fig:s2} and Table~\ref{tab:1} (see Appendix), showing excellent agreement with the calculations of \cite{2021MNRASGentileFusillo}\footnote{However, we find a significantly hotter temperature than \citet{Vincent2024A&A...682A...5V}, attributable to the higher resolution of our Keck/LRIS data over Gaia XP spectra.} and \cite{cristea2025halfringionizedcircumstellarmaterial}.


Given the position of RSG5-WD in the CMD, our model calculations indicate that it is a WD with a mass of approximately 1.05$\pm0.08$ M$_{\odot}$ (see Fig.~\ref{fig:1}). According to single-star evolutionary models, the progenitor mass of RSG5-WD is at maximum 6.8 M$_{\odot}$ \citep{Cummings2018ApJ...866...21C} or 7.0 M$_{\odot}$ \citep{Limongi2024ApJS..270...29L}, with corresponding progenitor evolutionary timescales not less than 57 Myr and 54 Myr respectively. Single-star evolution further suggests RSG5-WD requires $\sim$60 Myr of cooling time (see Fig.~\ref{fig:1}).

We assess whether RSG5-WD could form through single-star evolution within the cluster's age range using different stellar models. We simulated 1,000 realizations of RSG5-WD's photometry (including Gaia errors). For each, we derived the total age using \texttt{WD\_models}\footnote{\url{https://github.com/SihaoCheng/WD\_models}}, which incorporates MIST models \citep{Choi2016ApJ...823..102C} for pre-WD evolution. Comparing these ages to the cluster's range (22--57 Myr), we calculated the probability of the WD forming outside this interval. For 22 Myr, the probability that the WD could not have formed within the cluster age range was greater than 0.99, and for 57 Myr, the corresponding result was 0.976. Given that this result is model-dependent, we also repeated the analysis with other stellar evolution models. Using the \texttt{wdwarfdate}\footnote{\url{https://github.com/rkiman/wdwarfdate}} code \citep{Kiman2022AJ....164...62K} with different initial final mass relationships (IFMR) \citep{Marigo2020NatAs...4.1102M, Cummings2018ApJ...866...21C} (see Appendix Fig.~\ref{fig:sage}), we rule out ($p>0.99$) single-star formation of RSG5-WD within the cluster's maximum literature age (57 Myr; \citealt{Bouma2022AJ....164..215B}).


\subsection{Rotation and surface magnetic field}
\begin{figure*} 
	\centering
	\includegraphics[width=0.49\textwidth]{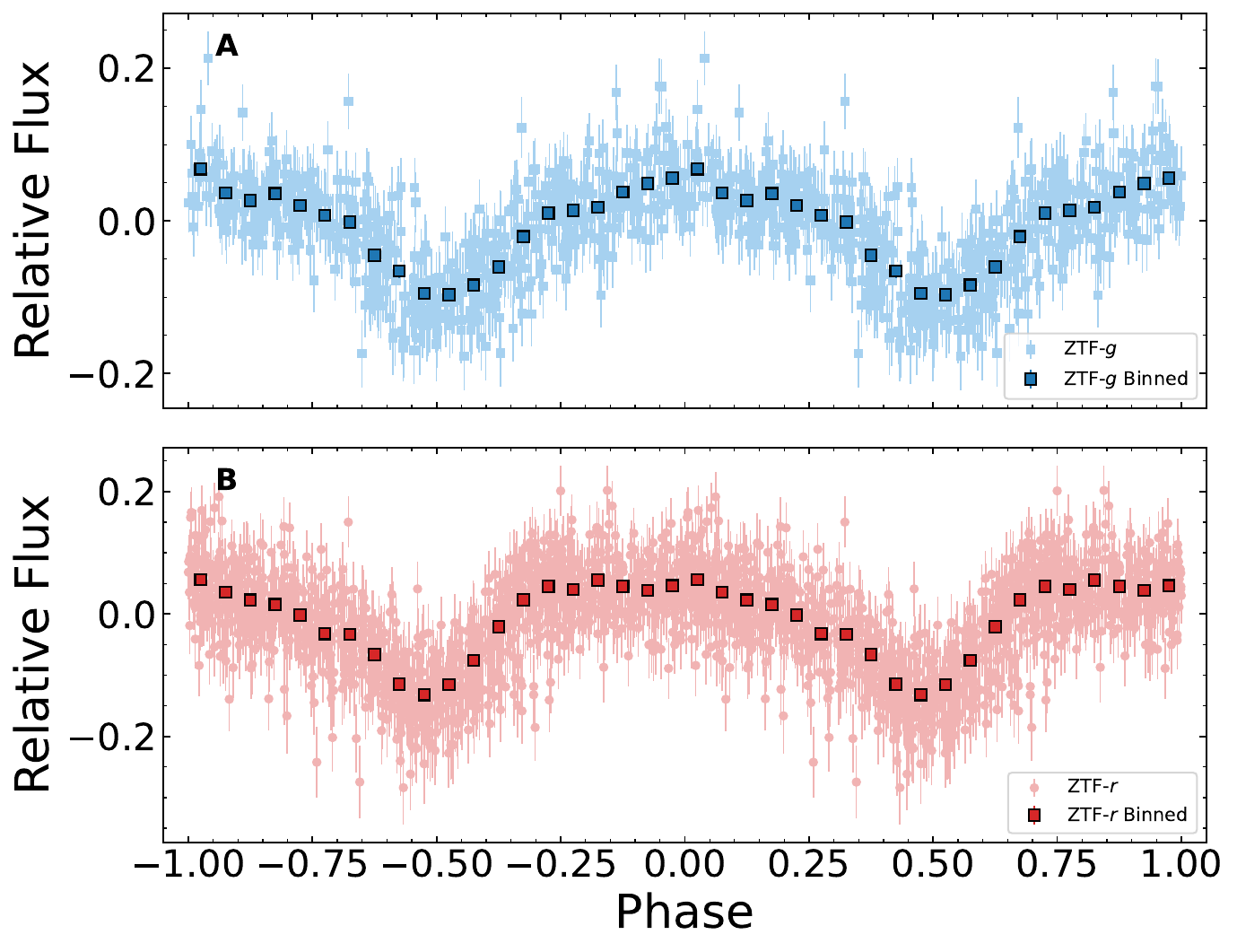} 
    \includegraphics[width=0.49\textwidth]{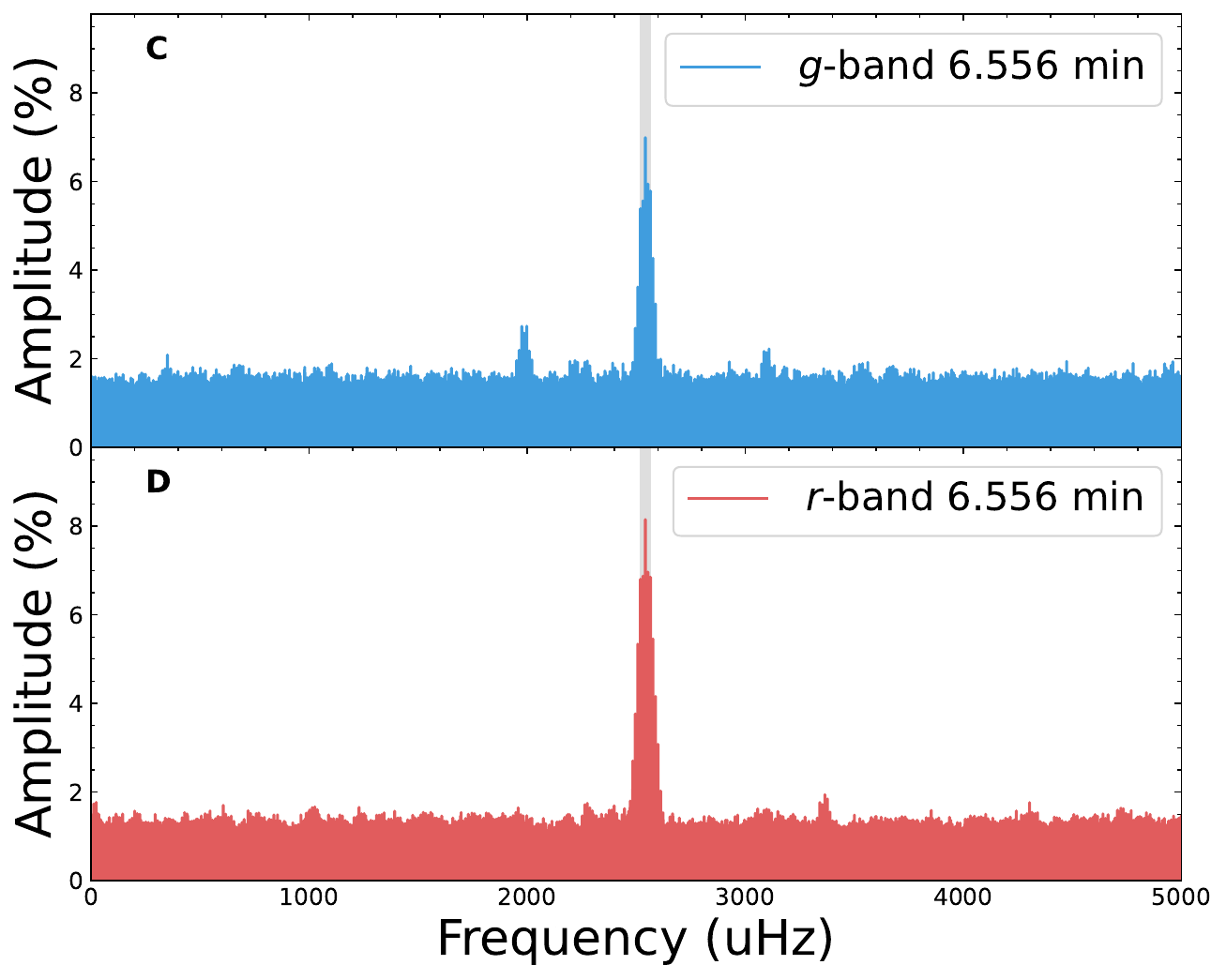}
    \caption{RSG5-WD light curve and power spectrum. (\textbf{A}): ZTF g-band light curves folded with a period of 6.556 minutes. The square markers represent binned data, where every 20 points are averaged into a bin. (\textbf{B}):Same as (\textbf{A}), but for r-band light curves. (\textbf{C}): The power spectrum in the g-band.  The gray shading highlights the frequency corresponding to the maximum value of the power spectrum.  (\textbf{D}): The power spectrum in the r-band.}
	\label{fig:2} 
\end{figure*}

\begin{figure*} 
	\centering
	\includegraphics[width=0.47\textwidth]{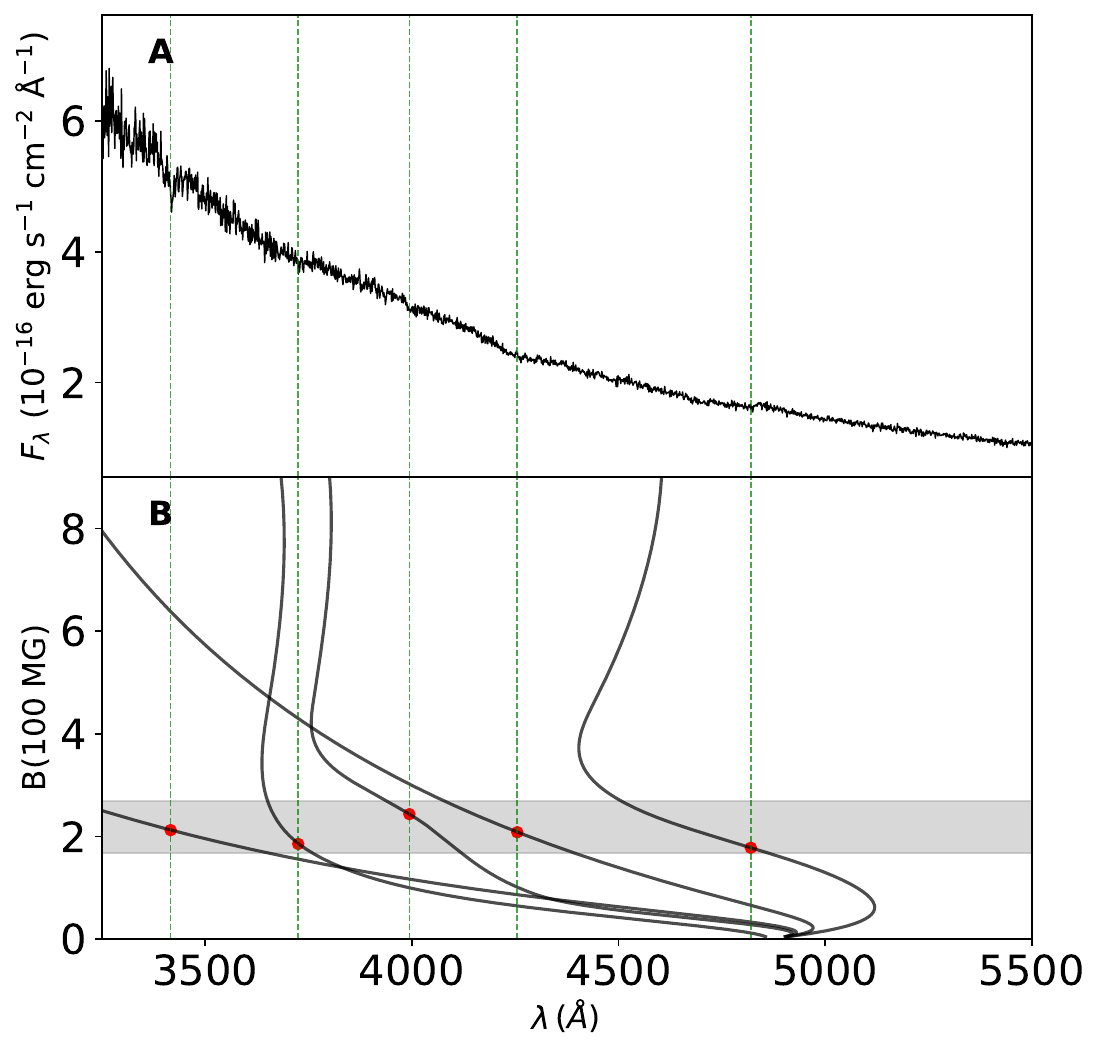} 
    \includegraphics[width=0.47\textwidth]{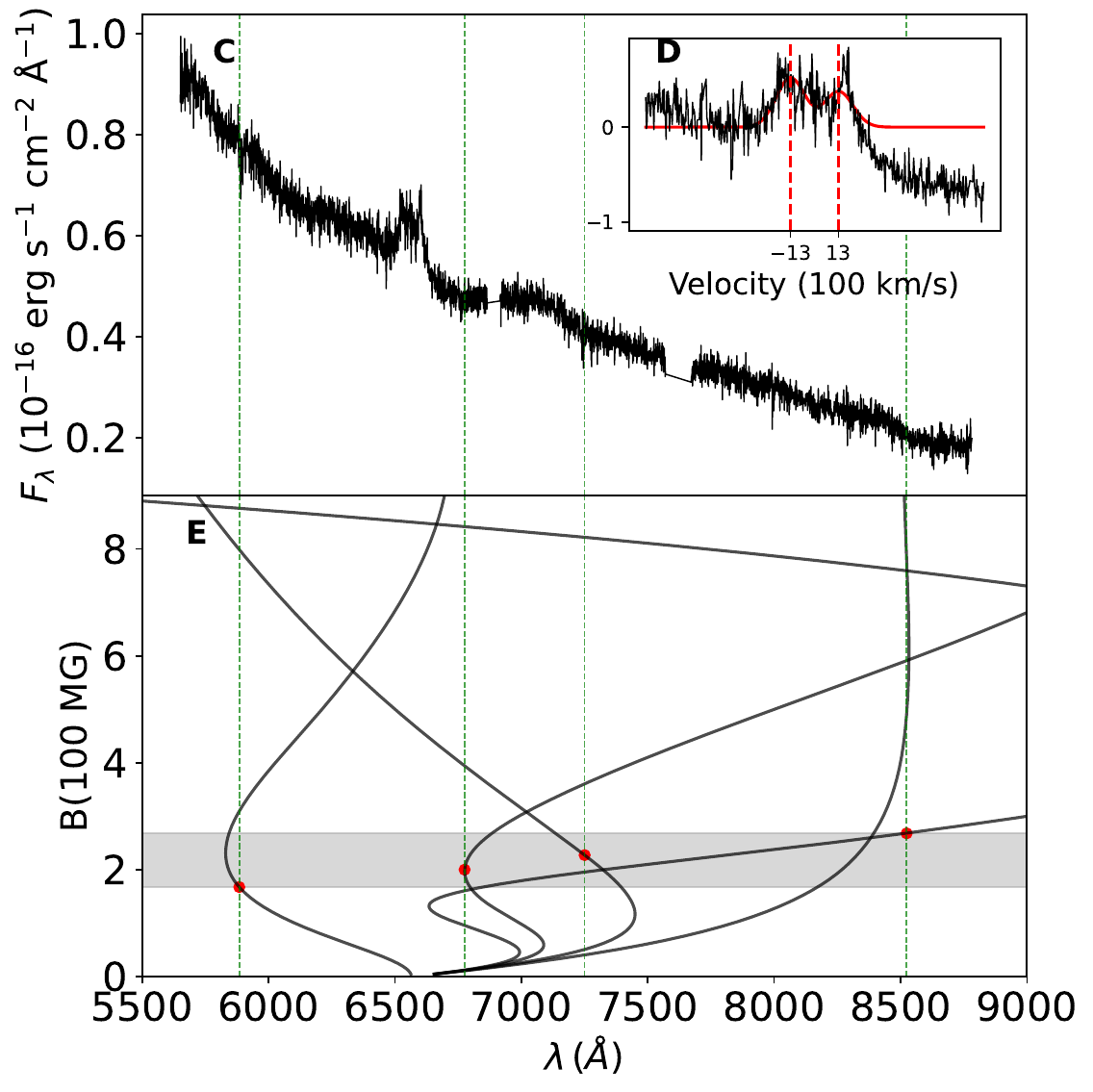}
    \caption{RSG5-WD spectrum with Zeeman components of Hydrogen indicated. (\textbf{A}):The blue section spectrum of RSG5-WD from Keck I telescope. (\textbf{B}): The wavelength as a function of magnetic field for the hydrogen lines. The green line represents the hydrogen absorption lines in the spectrum, and the shaded area indicates the corresponding magnetic field range. (\textbf{C}): Same as (\textbf{A}), but for red section spectrum. (\textbf{D}): The H$\alpha$ emission line range. The red solid line represents a double Gaussian fit to the double-peaked H$\alpha$ emission line, while the dashed lines indicate the corresponding peak positions. (\textbf{E}): Same as (\textbf{B}), but for red section spectrum. The average fit magnetic field strength is about 210 MG.}
	\label{fig:3} 
\end{figure*}

To determine if RSG5-WD exhibits any periodic variability, we obtained time-series photometric data for RSG5-WD in the Zwicky Transient Facility (ZTF) DR23\footnote{\url{https://irsa.ipac.caltech.edu/cgi-bin/Gator/nph-dd}} \citep{Bellm2019PASP..131a8002B}. Using the Lomb-Scargle periodogram, we detect a significant 6.556-minute periodic variability in RSG5-WD's light curves with an amplitude of $\sim$10-15\% (Fig.~\ref{fig:2}), consistently seen in both $g$- and $r$-band observations.
Fig.~\ref{fig:3} shows the Keck LRIS spectrum from Program ID C367.
We found that the spectrum of RSG5-WD shows no prominent features, with no significant absorption occurring at the wavelengths of the notable Balmer lines or helium lines. We have also not detected any contribution from a companion star in the spectrum of RSG5-WD, suggesting that it may be a single star or that its companion is nearly invisible compared to RSG5-WD. We found weak absorption features in the wavelength range of 3000\ \AA \ to 9000\ \AA, with the most notable absorption occurring at around $\sim$3400\ \AA \,(see the panel (\textbf{A}) of Fig.~\ref{fig:3}). We identified these weak absorption lines as Zeeman components of hydrogen's Balmer lines under a strong magnetic field ($\geq$ 170 megagauss (MG)), the average magnetic field is about 210 MG. Based on the transition table of hydrogen absorption lines under different magnetic fields provided in \cite{Ruder1994}, we estimate the magnetic field range of the RSG5-WD, the method is same as \cite{Caiazzo2021Natur.595...39C}. 


In summary, from the spectrum and photometric parameters, we identified RSG5-WD as a DA WD, with a surface gravity as high as $\log{\mathrm{g}}$ = $8.68^{+0.16}_{-0.14}$ cm/s$^2$, a surface temperature of $T_{\mathrm{eff}}$=$31,622^{+5531}_{-3439}$ K and a stellar mass of 1.05$\pm0.08\,M_{\odot}$. We also found a weak emission feature with a double peak structure near the Balmer line H$\alpha$ ($\sim$6563 \AA), indicating the presence of a high-velocity ionized circumstellar structure surrounding RSG5-WD. The H$\alpha$ emission line shows a clear double-peaked structure, corresponding to a velocity of about $\sim$1300 km/s relative to the center of mass (see the panel (\textbf{D}) Fig.~\ref{fig:3}). We also observed RSG5-WD using the Gemini Multi-Object Spectrograph (GMOS) \citep{Hook2004PASP..116..425H} on the North Gemini Telescope and obtained low-resolution spectra covering a similar wavelength range. Although the spectra from North Gemini had a relatively low signal-to-noise ratio, we still detected clear emission lines near the H$\alpha$ central wavelength. According to \citep{cristea2025halfringionizedcircumstellarmaterial}, this double-peaked Balmer emission was attributed to a half-ring of ionized circumstellar materials, possibly a debris originating through a tidal disruption of a planetary object,  fallback accretion of gravitationally bound ejecta from a binary merger event, or a wind from the surface of the WD.

\subsection{Origin of the WD's variability}
We now discuss the origin of the short-period variability in RSG5-WD. First, we can rule out WD pulsations, as the typical pulsation temperatures for DA WDs range from 10,500 to 12,500 K \citep{VanGrootel2012A&A...539A..87V} (see the dashed area in the right panel (\textbf{F}) of Fig.~\ref{fig:1}), while the temperature of RSG5-WD is far above this range. Additionally, its magnetic field may suppress gravity-mode (g-mode) pulsations \citep{Fuller2015Sci...350..423F, Loi2020MNRAS.493.5726L}, making the pulsation unlikely to reach about 15\% (see Fig.~\ref{fig:2}). We can also rule out the possibility of binary eclipses, as the light curve of RSG5-WD is very smooth. If the variability was due to eclipses, RSG5-WD would have been stretched by its companion, creating an ellipsoidal signal. Our calculations show that in a 6.556 minute orbit, a typical main-sequence star or a low-mass red dwarf would be completely disrupted by the WD in such a short orbit, while denser objects like neutron stars or black holes would not be able to stretch RSG5-WD this much. 


The only viable explanation is that the observed short-period variability is the result of the WD's rotation. The strong magnetic field can cause changes in variations in the continuum opacities and surface temperature. As the star rotates, we can detect variations in the flux caused by the changing magnetic field strength across different parts of the star's surface, which can lead to a light curve variation of close to 10\% \citep{Ferrario1997MNRAS.292..205F, Ferrario2015SSRv..191..111F, Hernandez2024MNRAS.528.6056H, Pelisoli2023NatAs...7..931P}. Considering the presence of a circumstellar disc that could obscure light, it should not be challenging to achieve light variations exceeding 10\%. We also analyzed phase-resolved spectra obtained from LRIS/Keck I (Program ID C266, see Fig.~\ref{fig:s4}). The relative faintness of RSG5-WD resulted in high noise levels in individual spectral observations, which made it challenging to extract spectral details as we did with the phase-averaged spectra. Nevertheless, we are still able to clearly observe significant variations in spectral flux of approximately 15\% on a timescale of just a few minutes.

\section{Formation pathway} \label{sec:floats}
The single-star formation timescale for RSG5-WD significantly exceeds the cluster age. Together with its rapid rotation and strong magnetic field ($\sim\!2\times10^8$ G), this conclusively points to a binary origin. Therefore, it is reasonable to infer that it must have recently undergone a binary merger event. Previous studies have shown that recently merged WD could have a strong magnetic field and high rotational speed \citep{Tout2008MNRAS.387..897T, Caiazzo2021Natur.595...39C, Jewett2024ApJ...974...12J}. This is because of the strong magnetic dynamo that could arise during the merger, and it would be rapidly spinning due to the conservation of orbital angular momentum \citep{Schwab2021ApJ...906...53S}. To test the above hypothesis, we conducted extensive computer simulations of binary star evolution under a wide range of initial conditions \citep{Hurley2002MNRAS.329..897H}. These included varying the masses of the primary and secondary stars, the initial orbital periods (or orbital radii), and the eccentricities of their orbits. We evolved all binary systems for 40 Myr and searched for initial parameters that could produce a WD with a mass between 1.0 and 1.1 M$_{\odot}$ within this timescale. As expected, almost all WDs had undergone a merger with their companion stars. We also found that some binary systems produced a WD with an invisible neutron star or black hole. However, the resulting WDs in such systems often had masses either too large ($>1.3$ M$_{\odot}$) or too small ($<1.0$ M$_{\odot}$). In addition, the probability of forming a WD through this pathway is extremely low, ranging from 1\% to $10^{-5}$ of those formed through mergers. Our simulations show these WDs do not ultimately reside in binary systems. Therefore, we conclude that RSG5-WD must have experienced a merger event with a companion star, rather than being located near a neutron star or a black hole. 

Our simulations show that when the mass of the primary star is between 8.5 and 11 M$_{\odot}$, the mass of the companion star is between 0.1 and 7.5 M$_{\odot}$, and the initial orbital period of the binary system is between 1 and 1000 days, it is possible to produce the observed intermediate-mass WD within 35 Myr. Given that such massive stars often form in binary systems \citep{Duch2013ARA&A..51..269D}, this range of initial conditions is not restrictive. To produce a WD with a mass close to the observed value (1.04 M$_{\odot}$), the optimal initial conditions involve a combination of an intermediate mass star (approximately 9-10 M$_{\odot}$) and a low mass star (less than 2.5 M$_{\odot}$), with an orbital period between 150 and 1000 days. 

\begin{figure*} 
	\centering
	\includegraphics[width=0.46\textwidth]{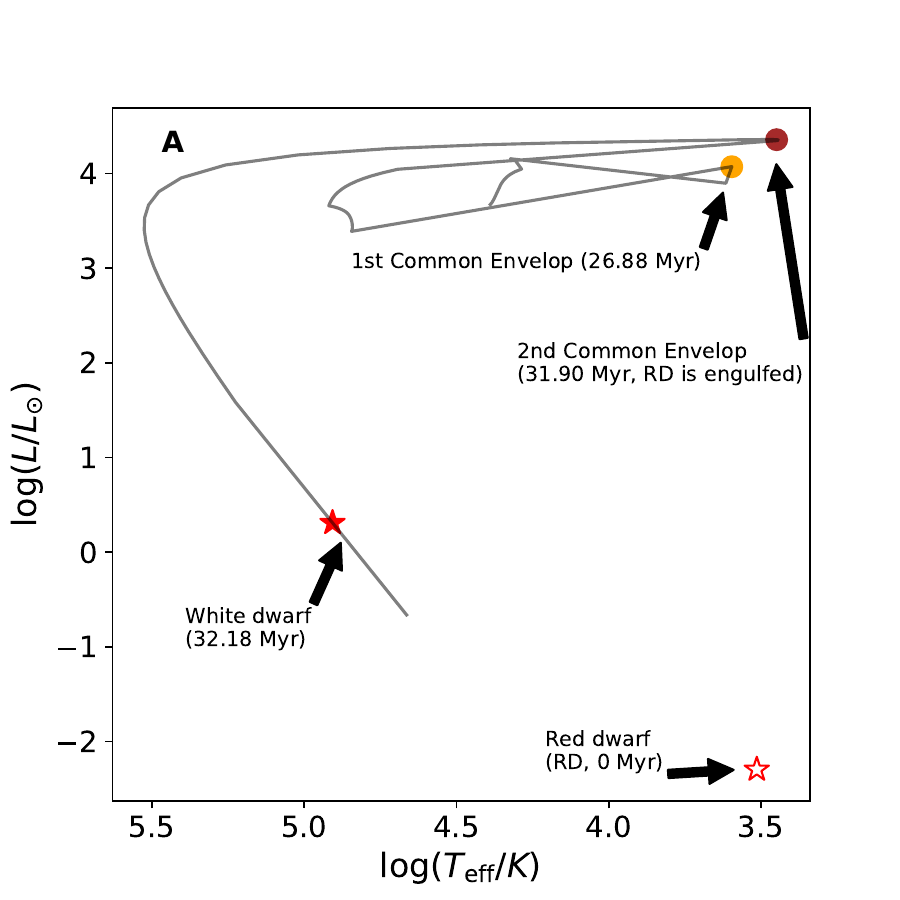}
    \includegraphics[width=0.42\textwidth]{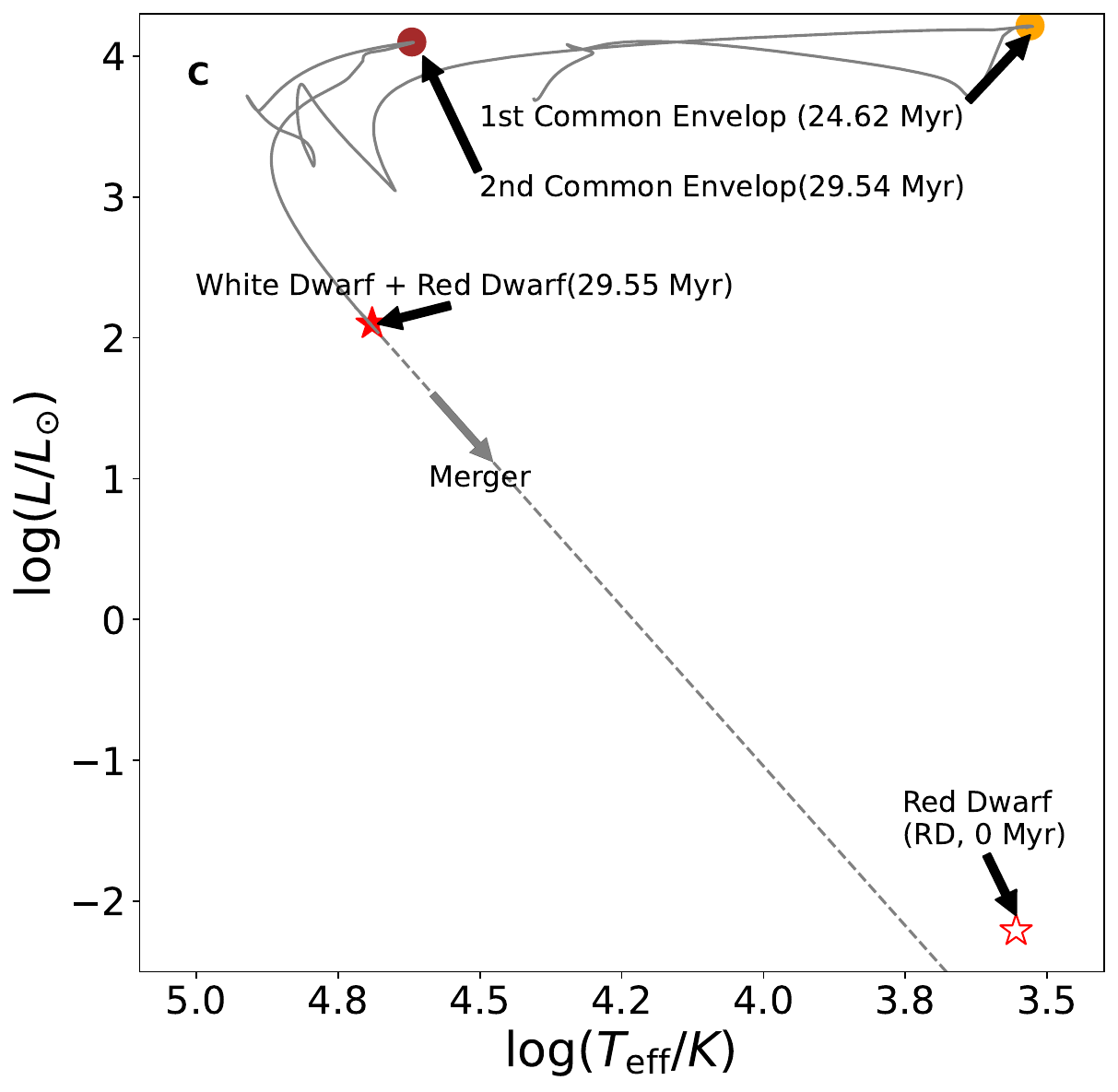}\\
    
    \includegraphics[width=0.46\textwidth]{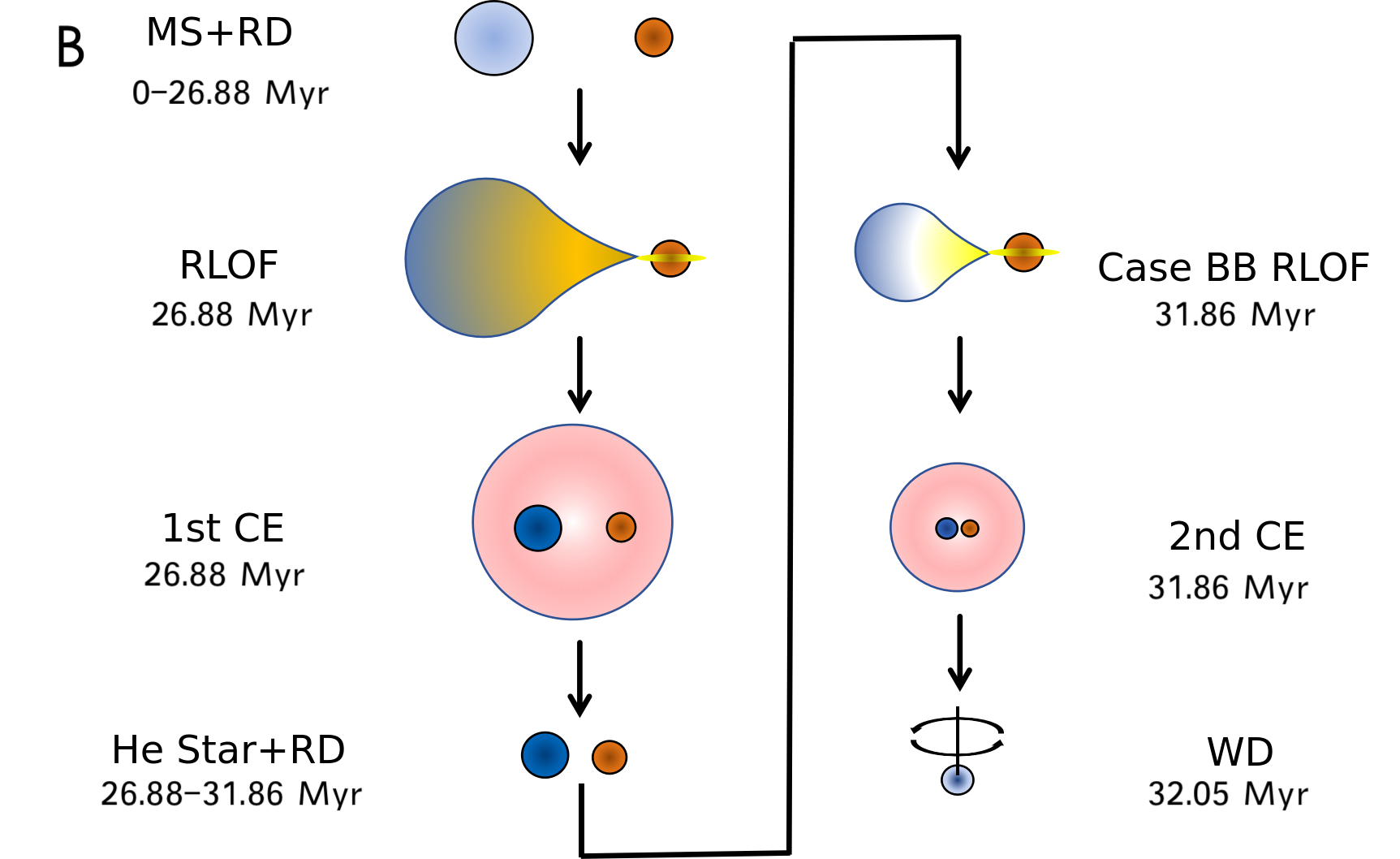}
    \includegraphics[width=0.46\textwidth]{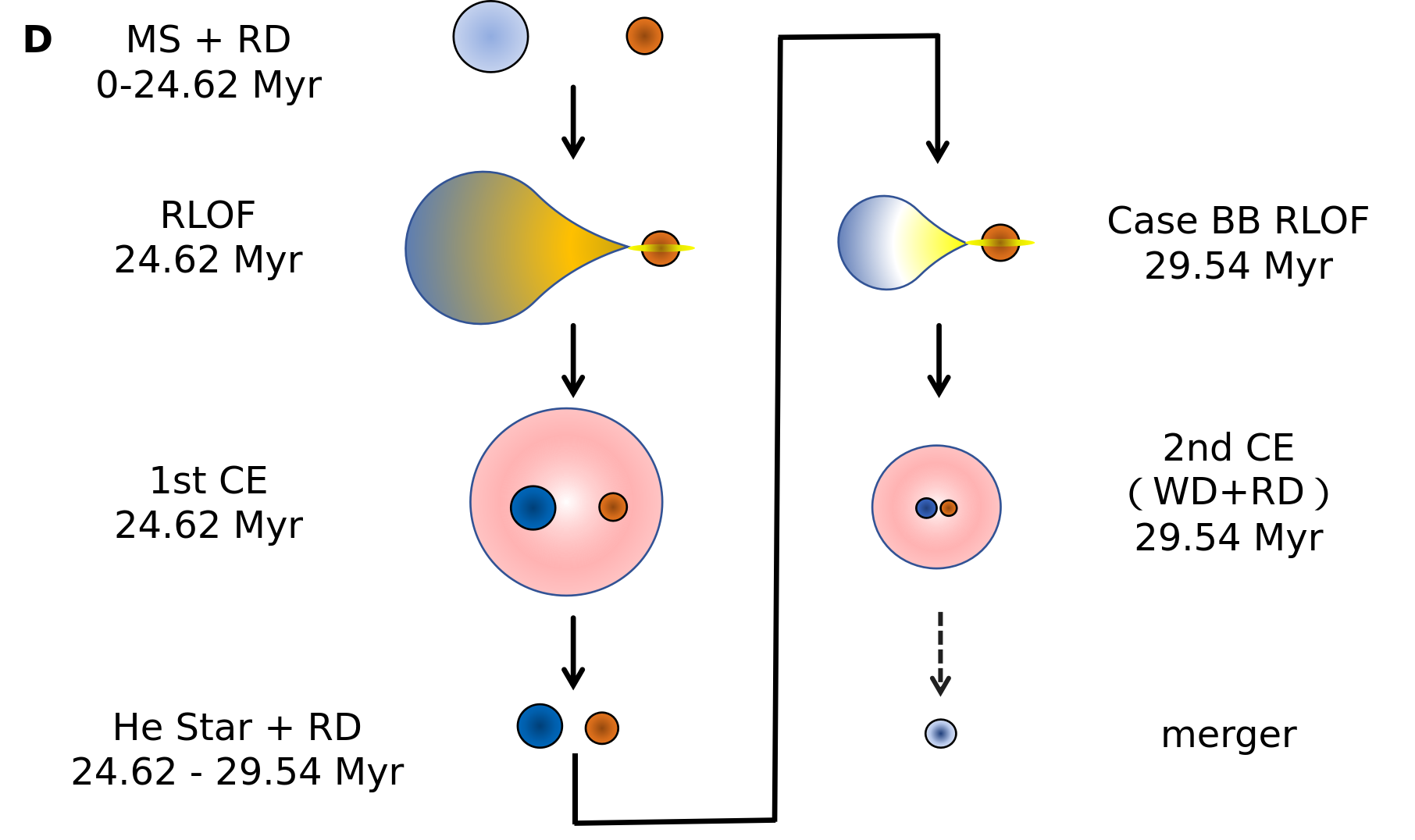}
    \caption{An example illustrating how binary interaction can result in the formation of the observed white dwarf within 35 Myr. (\textbf{A}): The evolution of the binary star on the Hertzsprung-Russell diagram from BSE result. (\textbf{B}): The evolutionary flow chart according to BSE result. In flow chart, the full name of the corresponding abbreviation: MS + RD: Main Sequence + Red Dwarf; RLOF: Roche Lobe Overflow; 1st CE: First Common Envelope Phase; He Star + RD: Helium Star + Red Dwarf; Case BB RLOF: Case BB Roche Lobe Overflow (The mass loss of a helium star due to Roche lobe overflow); 2nd CE: Second Common Envelope Phase; WD: White Dwarf. (\textbf{C}): The evolution of the binary star on the Hertzsprung-Russell diagram from MESA. (terminated at WD+RD stage due to MESA’s merger resolution limits). (\textbf{D}): The evolutionary flow chart. In flow chart, the full name of the corresponding abbreviation is same as pannel (\textbf{B})}
	\label{fig:4} 
\end{figure*}

Here, we show an example, with initially a progenitor binary system containing a 9.5 M$_{\odot}$ star and a 0.2 M$_{\odot}$ red dwarf (see Fig.~\ref{fig:4}). The initial separation between two binary components is about 400 times the solar radius ($\sim2.8\times10^8$ km. or an initial orbital period of about 300 days.). At around 26.8 Myr, the hydrogen in the core of the primary star is nearly exhausted, causing it to expand rapidly and evolve into a giant star. This triggers the first unstable mass transfer, rapidly forming a common envelope (CE). In the meantime, the binary system's orbit rapidly decays while releasing orbital energy that ejects the CE. As a result, a naked helium star with a mass of about 2 M$_{\odot}$ is formed, with a red dwarf that has not yet had a significant increase in mass, and the binary orbital separation is reduced to $\sim$1.95 $R_{\odot}$ (equals to an orbital period of only 2.88 hours). After approximately 3 Myr, the helium star evolves again, leading to the system undergoes a second unstable mass-transfer episode, producing a CE again. The binary orbit subsequently shrank, leading to stellar merger. After ejecting the common envelope, the system ultimately produced a $1.03\,M_\odot$ WD at $\sim\!32$\,Myr. 


Since BSE is a rapid but simplified population synthesis model, we verified its results using high-precision \texttt{MESA} stellar evolution code \citep{Paxton2011ApJS..192....3P, Paxton2013ApJS..208....4P, Paxton2015ApJS..220...15P, Paxton2018ApJS..234...34P, Paxton2019ApJS..243...10P}. Using the same initial parameters as BSE, \texttt{MESA} produces consistent evolutionary phases with timing variations $<2$\,Myr at critical stages. The main discrepancy are: (1) the second unstable mass-transfer produces a less prominent CE compared to the first episode; and (2) the primary star has already evolved into a $\sim\!1.0\,M_\odot$ WD during this phase. We highlight that current \texttt{MESA} versions cannot model post-merger evolution, limiting our simulation to the pre-merger phase (\texttt{MESA} terminates when the CE orbital period falls below 5 minutes). Nevertheless, even if the merger accretes the companion's entire $0.2\,M_\odot$ envelope, the final product remains a WD. Fig.~\ref{fig:4} shows the binary system's evolution on an HR diagram, alongside a schematic illustrating these processes, for both BSE and \texttt{MESA} calculations.

We caution that stellar evolution uncertainties mean other pathways may exist beyond our models. At least, our calculations, combined with the age constraint of RSG5-WD derived from its host star cluster, provide further evidence that it likely experienced binary interaction in the past.

The binary merger model provides a good explanation for the observed H$\alpha$ emission. Research has already shown that when the mass of the companion star within the CE is sufficiently small, the CE does not immediately eject \citep{Nordhaus2011PNAS..108.3135N}. Instead, the companion star may be tidally disrupted by the strong tidal forces as it spirals in towards the primary star (a pre-WD), potentially leaving behind a remnant around the WD that forms a circumstellar disc. The circumstellar disc also possesses turbulence and shear, which can amplify magnetic fields through dynamo action  \citep{Nordhaus2011PNAS..108.3135N}. The rotational speed of the half-ring of ionized circumstellar debris depends on its distance from the WD. Assuming that the centripetal force of the half-ring of ionized circumstellar debris is provided by the gravitational force of the WD, it would need to be located approximately 13 $R_{\odot}$ (about 82,500 km)
from the WD to achieve a motion speed of $\sim$1300 km/s for materials in the half-ring. At this point, the orbital period of the disc around the star is about 6.6 minutes, which is exactly equal to the rotation period of the WD. This indicates that the half-ring of ionized circumstellar debris may be locked by the star's strong magnetic field.

\section{Discussion}
Recently, \cite{cristea2025halfringionizedcircumstellarmaterial} conducted a comprehensive multi-wavelength study of RSG5-WD, combining UV-to-IR imaging with high-S/N spectroscopy. Most of our results are consistent. Their spectral energy distribution (SED) fitting yields $M = 1.12 \pm 0.03\,M_{\odot}$ and $T_{\rm eff} = 35,\!500 \pm 300$~K. Their time-domain photometry in UV/optical bands independently confirms the 6.6-minute variability period of RSG5-WD. They reach the same conclusion that the 6.6-min photometric period reflects the rotational period of the WD. Consistent with our models, their analysis combining PARSEC isochrones and cooling tracks \citep{Bedard2020ApJ...901...93B} shows RSG5-WD requires $70+60\pm10$ Myr of evolution as a single star, significantly exceeding their estimated cluster's age ($\sim$45 Myr).They similarly conclude RSG5-WD is a highly magnetic WD, though with a stronger surface field (400--600 MG) than our estimate ($\sim$200 MG). Given their multi-wavelength, high-S/N spectroscopic data, we consider their measurement more reliable, suggesting our value may represent only a lower-limit estimate.

A key advance in the work by \cite{cristea2025halfringionizedcircumstellarmaterial} is their time-resolved spectroscopy of RSG5-WD, which enables direct investigation of the H$\alpha$ emission-line structure.
One of their key conclusion identifies the double-peaked H$\alpha$ emission as originating from a half-ring of ionized circumstellar debris that shares the WD's 6.6-minute rotation period. Their time-resolved spectroscopy measured the half-ring of ionized circumstellar debris's mean RV, which matches the cluster's stellar RV distribution very well (see Section \ref{sec:style}). Since the half-ring of ionized circumstellar debris is magnetically locked to the WD (forming a rigid co-rotating system), this kinematic agreement provides independent evidence for RSG5-WD's cluster membership. Their SED fitting excludes any red dwarf companion around RSG5-WD. This rules out continuous accretion from a binary partner, indicating the half-ring of ionized circumstellar structure likely formed from post-merger debris, or formed through the stellar wind from the WD surface. Based on the spatial and kinematic association of RSG5-WD with the cluster, combined with the results from our modeling calculations, we favor the former scenario: the half-ring of ionized circumstellar debris is the remnant of a binary merger event. 

Although \cite{cristea2025halfringionizedcircumstellarmaterial} acknowledge RSG5-WD's astrometric alignment with cluster RSG5, they argue that even an 8$M_\odot$ progenitor (the theoretical single-star upper limit) would require $\sim$40 Myr to evolve into a WD - fundamentally incompatible with the cluster's age. They therefore conclude RSG5-WD cannot be a cluster member, representing the key divergence from our interpretation. We suggest that an alternative perspective might consider the potential contribution of binary evolution pathways, which could permit WD progenitors to exceed the nominal single-star mass limit. Our models indicate that WDs forming within 40 Myr systematically require progenitor masses $\geq$8.5\,$M_{\odot}$ --- a mass regime that falls just beyond the upper boundary examined in \cite{cristea2025halfringionizedcircumstellarmaterial}. Notably, binary interactions may facilitate the more rapid ($\lesssim$40~Myr) evolution of such massive ($\geq$8.5\,$M_{\odot}$) stars into WDs compared to purely single-star evolutionary timescales.

\cite{cristea2025halfringionizedcircumstellarmaterial} also attribute RSG5-WD to binary merger formation, they specifically propose a double WD merger scenario. However, given RSG5-WD shows good 6D kinematic agreement (position, parallax, proper motion, and RV) with cluster RSG5, providing compelling evidence for its cluster membership. According to our model calculations, we propose that RSG5-WD originated from a non-degenerate binary merger rather than a double WD merger, as the cluster's age is insufficient to produce two WDs. Our interpretation is supported by theoretical simulations \citep{Tout2008MNRAS.387..897T, Briggs2015MNRAS.447.1713B, Nordhaus2011PNAS..108.3135N}, as well.


\section{Conclusion}
The presence of RSG5-WD with a total age of about $\sim$35 Myr indicates that we may have detected an extremely young WD formed through binary-channel. Its membership in cluster RSG~5 enables precise age determination, revealing detailed formation history and evolutionary pathways.  These results suggest binary interactions could enable WDs to form earlier than standard age limits permit, providing a unique observational probe of this phenomenon. WD binary merger events are important sources of gravitational waves in the universe, which will be detected by next-generation space-based gravitational wave observatories \citep{Mei2021PTEP.2021eA107M,Lau2025PhRvD.111b4039L}. Since most massive stars in young clusters exist in binaries, binary interactions are expected to be frequent. Other young massive clusters might have also produced WDs like RSG5-WD, which can help us refine our understanding of the conditions under which binary stars form within clusters. Star clusters could serve as an ideal environment for the formation of such strongly magnetized compact objects.

\begin{acknowledgments}
H.Y., J.L., L.W., and C.L. acknowledge support from the Natural Science Foundation of China (NSFC) through grant 12233010. H.G. acknowledges the support from the Strategic Priority Research Program of the Chinese Academy of Sciences (grant Nos. XDB1160201), the NSFC (Nos. 12288102, 12090043,12173081), the National Key R\&D Program of China (No. 2021YFA1600403), Yunnan Fundamental Research Projects (Nos. 202401BC070007, and the International Centre of Supernovae, Yunnan Key Laboratory (No. 202302AN360001). Z.G. is funded by ANID, Millennium Science Initiative, AIM23-001. Z.G. is supported by the China-Chile Joint Research Fund (CCJRF No.2301) and Chinese Academy of Sci
ences South America Center for Astronomy (CASSACA) Key Research Project E52H540301. We thank Dr. Zach Vanderbosch from the California Institute of Technology for the observations of RSG5-WD using the Keck telescope, and this paper utilizes the observational data he provided as the PI. We thank Ilaria Caiazzo from the Institute of Science and Technology Austria for her scientific discussions and constructive suggestions on this work.
\end{acknowledgments}

\begin{contribution}
C.L. and H.G. are responsible as the corresponding authors for the entire work. C.L. conceived the study and designed the methodology. H.G. led the theoretical calculations and their physical interpretation. H.Y. is the main executor of this work; he first discovered the source of this study, wrote the main content of the article, and carried out all steps from data analysis to scientific calculations. J.L., R.Z. and L.W. contributed equally to this work; they jointly completed the main steps required, such as spectral analysis, magnetic field calculations and stellar evolution calculations. J.E. provided scientific insights and suggestions for the theoretical explanation of this work. G.L. fitted the spectra of the white dwarf and determined its parameters. L.R. analyzed and interpreted the light curve of the white dwarf. C.F. observed RSG5-WD using the Keck telescope. Z.C., C.C., B.M., and S.X. conducted observations of the white dwarf in this work (using the GMOS/North-Gemini) and offered comments and suggestions for the article's writing. 
C.F., 
Y.S., Z.L., and X.Z. contributed scientific discussions for the section on binary star evolution in this work.



\end{contribution}

%
\facilities{Gaia (DR3), Gemini-North (GMOS), Keck:I (LRIS), Zwicky Transient Facility (DR 23)}



\appendix 


\begin{figure} 
	\centering
	\includegraphics[width=0.8\textwidth]{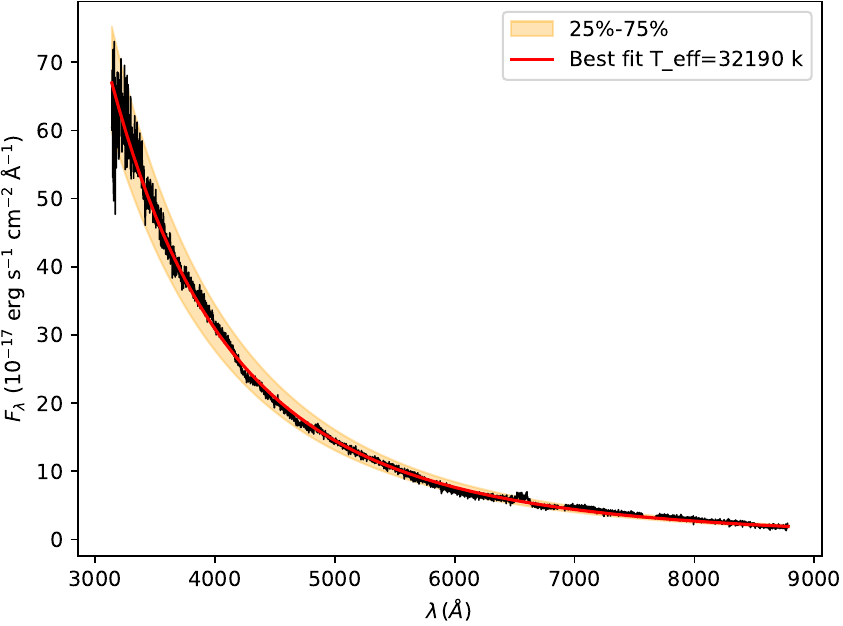} 
	\caption{Phase-averaged spectrum with the best-fitting blackbody model. The black line represents the spectra of RSG5-WD from Keck, the red line indicates the best-fit result, and the yellow shaded area represents the 25\% to 75\% confidence interval based on the MCMC fitting.} 
	\label{fig:s1} 
\end{figure} 

\begin{figure} 
	\centering
	\includegraphics[width=0.8\textwidth]{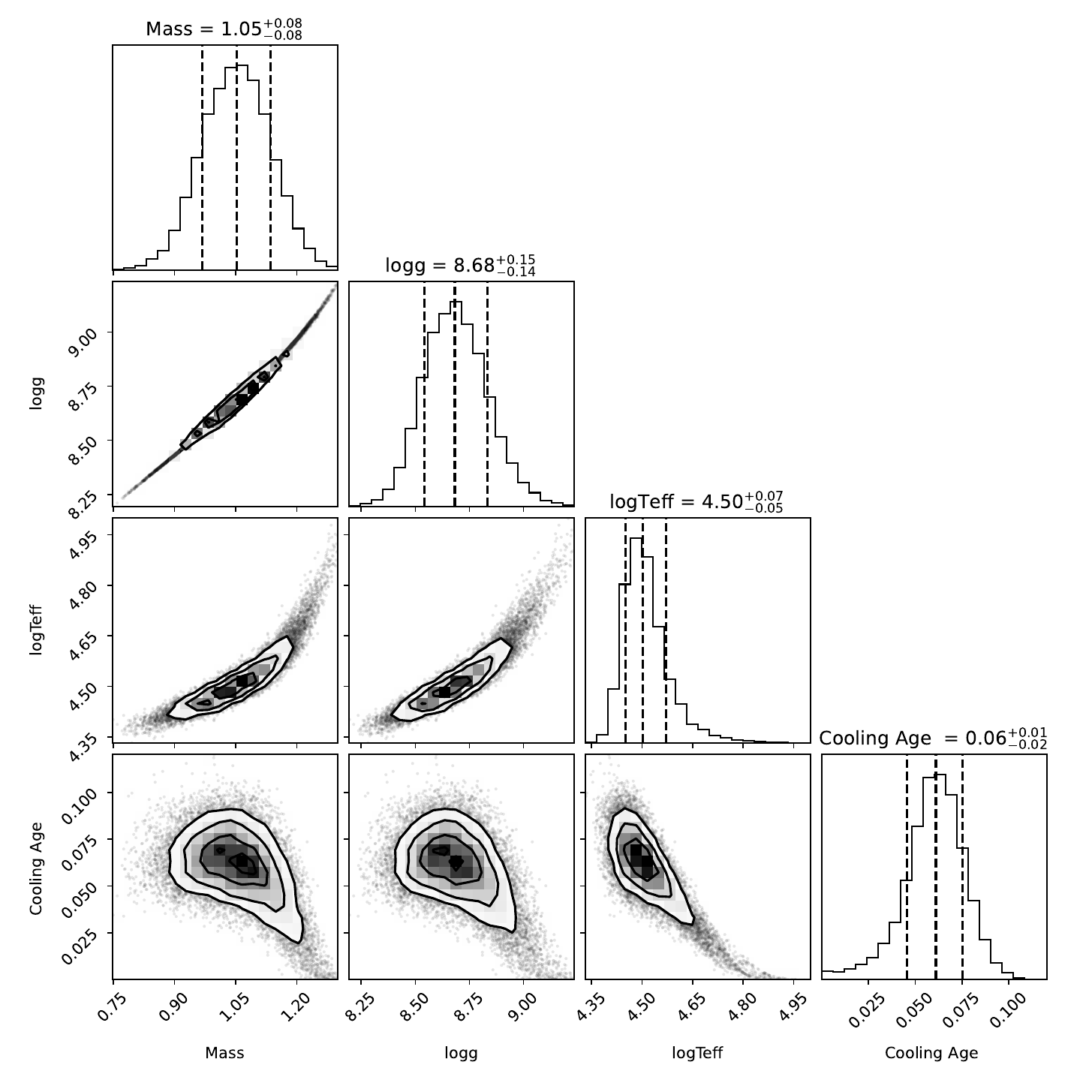} %
	\caption{Corner plots. Corner plots for parameter estimates based on Gaia photometry.}
	\label{fig:s2}
\end{figure}

\begin{figure}[h]
    \centering
    \includegraphics[width=0.8\linewidth]{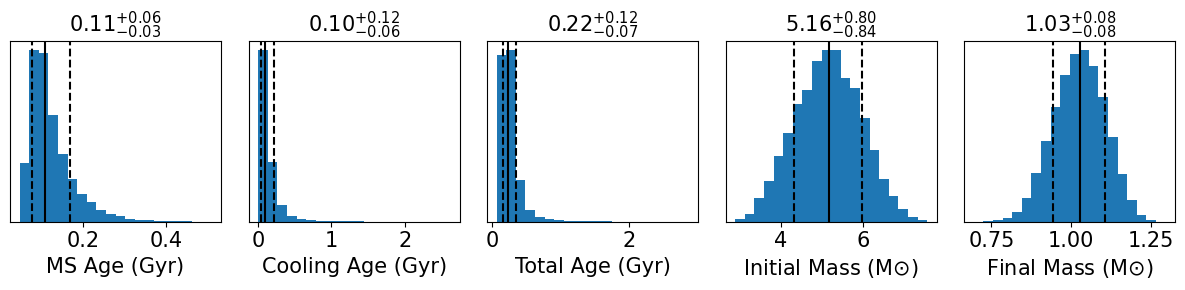}
    \includegraphics[width=0.8\linewidth]{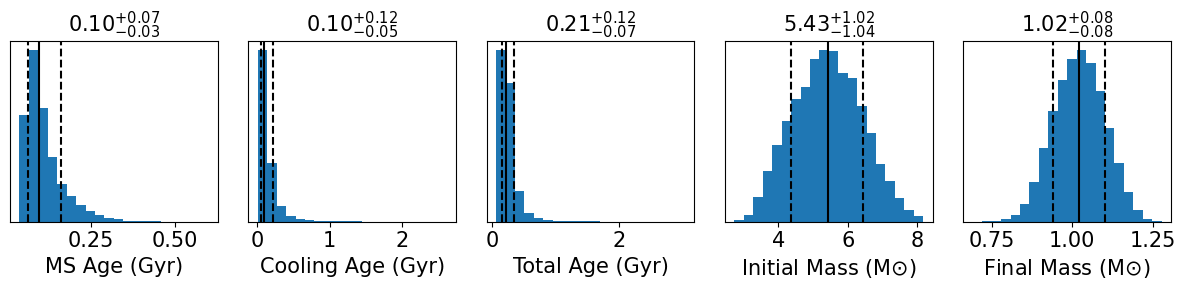}
    \includegraphics[width=0.8\linewidth]{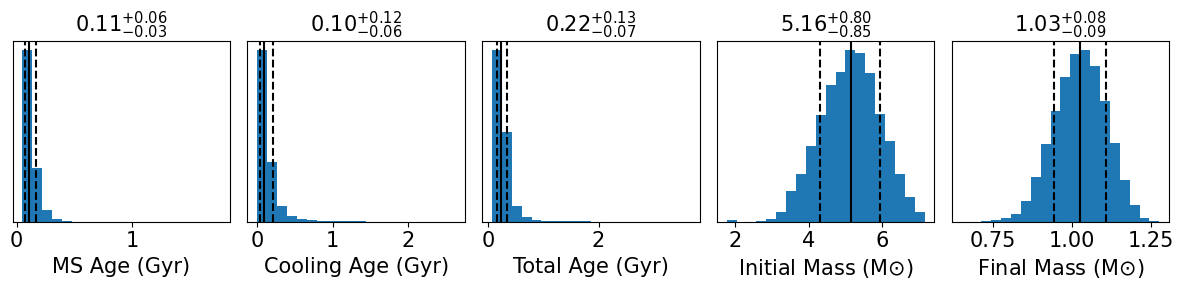}
    \caption{
        Comparison of RSG5-WD system ages and physical parameters under three different IFMR:
        (Upper panel) The MIST IFMR from \cite{Cummings2018ApJ...866...21C},
        (Middle panel) The PARSEC IFMR from \cite{Cummings2018ApJ...866...21C}, 
        (Lower panel) The IFMR from \cite{Marigo2020NatAs...4.1102M}.
        The physical parameters include main sequence age (MS age), cooling age, total age, initial mass, and final mass.
    }
    \label{fig:sage}
\end{figure}

\begin{table}[ht]
  \centering
  \caption{The parameters of RSG 5 cluster and RSG5-WD.}
  \label{tab:1}
  \begin{threeparttable}
    \begin{tabular}{|c|c|c|c|c|c|}
      \hline
      Parameter & RSG 5 & RSG5-WD & RSG5-WD\tnote{a} &RSG5-WD\tnote{b} & RSG5-WD\tnote{c}\\
      \hline
      $\alpha\ (\deg)$               & $303.909$            & $302.137$           & --- &---&---\\
      $\beta\ (\deg)$                & $45.738$             & $44.827$            & --- &---&---\\
      $\mu_\alpha\ (\mathrm{mas/yr})$& $3.645\pm0.02$       & $3.217\pm0.15$      & --- &---&---\\
      $\mu_\delta\ (\mathrm{mas/yr})$& $1.597\pm0.02$       & $1.328\pm0.16$      & --- &---&---\\
      $\pi\ (\mathrm{mas})$          & $2.967\pm0.004$      & $2.915\pm0.13$      & --- &---&---\\
      Mass\ ($M_{\odot}$)            & $138.401\pm15.80$    & $1.05\pm0.08$       & $1.02$ & $1.02$&$1.12\pm0.03$\\
      $\log g\ (\mathrm{cm/s^2})$    & ---                  & $8.68^{+0.16}_{-0.14}$ & $8.62$ & $8.62$&$8.8$\\
      $\log T_{\mathrm{eff}}\ (K)$   & ---                  & $4.50^{+0.07}_{-0.05}$ & $3.46$ & $4.47$& $4.55^{+0.04}_{-0.04}$\\
      \hline
    \end{tabular}
    \begin{tablenotes}
      \footnotesize
      \item[a] Parameters estimated by \citet{Vincent2024A&A...682A...5V}.
      \item[b] Parameters estimated by \citet{2021MNRASGentileFusillo}.
      \item[c] Parameters estimated by \citet{cristea2025halfringionizedcircumstellarmaterial}.
    \end{tablenotes}
  \end{threeparttable}
\end{table}



        
    

\begin{figure}[h]
    \centering
    \includegraphics[width=0.8\textwidth]{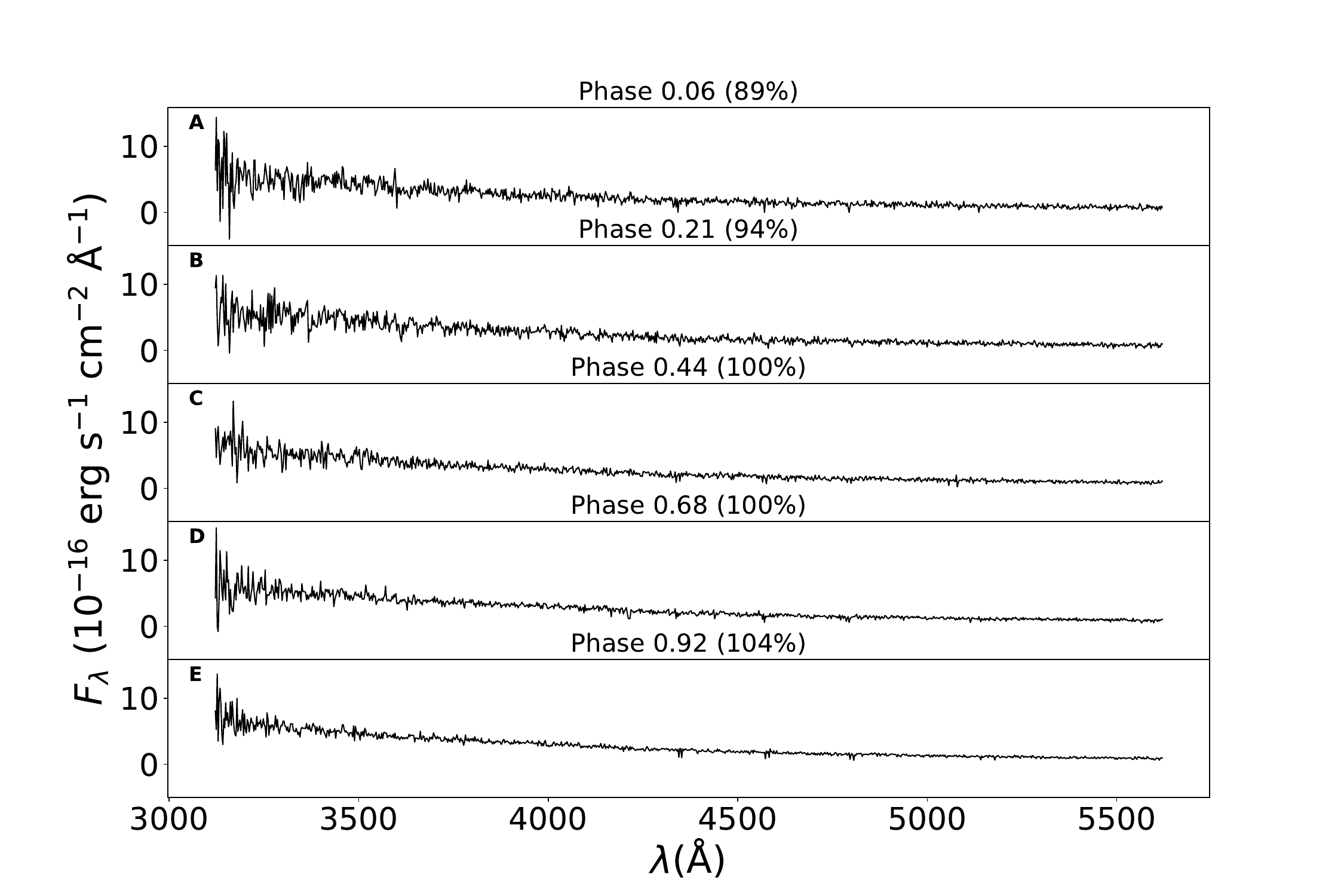}
    \includegraphics[width=0.8\textwidth]{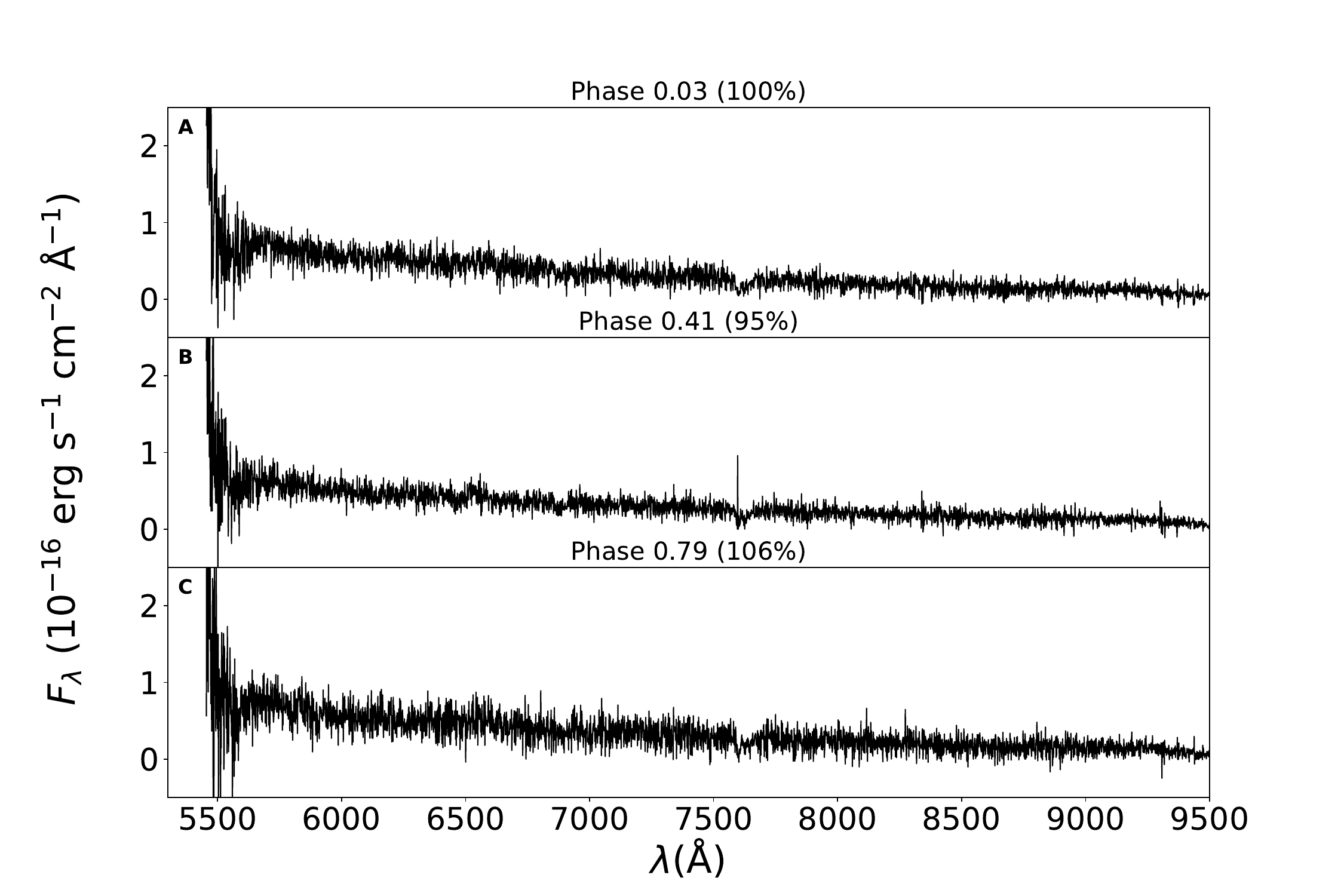}
    \caption{The phase-resolved spectra of RSG5-WD. (\textbf{A}): The blue-end spectra from program C266. We arranged the spectra in order of phases from top to bottom, and the percentage represents the relative flux ratio, where the reference spectrum is set to 100\%. (\textbf{B}): The red-end spectra program C266.}
    \label{fig:s4}
\end{figure}

\section{Identification of white dwarf members} \label{search method}

We performed an extensive matching process based on position and kinematics information. We expanded the positional selection range to $\theta < 5 \times r50$, where $\theta$ represents the angular distance of a WD from the OC center, and $r50$ is the half-number radius, as provided in the cluster parameters of \cite{Hunt2024A&A...686A..42H}. Next, for the parallax and kinematic matching, the filtering conditions are as follows:
\begin{equation}
\left( \pi, \mu_{\alpha}, \mu_{\delta} \right) < 1.6\times P 
\end{equation}
where $\pi$ represents the parallax, $\mu_{\alpha}$ and  $\mu_{\delta}$
represent the proper motions in right ascension and declination, and $P$ represents the parameter range of OC members.

We then applied the method described in \cite{ O'Grady2023ApJ...943...18O, Grondin2024arXiv240704775G} to improve the samples' reliability. 
This approach has been successful in removing foreground contamination from the Magellanic Clouds \citep{O'Grady2023ApJ...943...18O},  \cite{Grondin2024arXiv240704775G} applied it to identify WD main sequence (MS) binaries in OCs, achieving reliable results. We used the Gaia DR3 astrometric measurements for OC's stars to construct two separate covariance matrices ($\mathbf{C}$) for each cluster: one describes the 3D spatial distribution ($\vec{\mu} = [\alpha, \delta, \pi]$) and another describes the 2D kinematic distribution ($\vec{\mu} = [\mu_{\alpha}, \mu_{\delta}]$). 
Following \cite{Grondin2024arXiv240704775G}, we take measurement uncertainties into account by minimizing the total negative log-likelihood for a set of matrices, each weighted by the measurement uncertainties of a single object. This yields an optimal covariance matrix, $\mathbf{C_{*}}$, which is used in the rest of our analysis. 
Then, for each WD obtained from the initial screening, we calculate its $\chi^2$ value in the 2D kinematic and 3D spatial distributions: 
\begin{equation}
    \chi^2 = ({\mu} - {X})^T\mathbf{C}_{*}^{-1}({\mu} - {X})
\end{equation}
where ${X}$ represents the median spatial or kinematic properties of the OC's members. To determine if a WD is spatially or kinematically consistent with a specific cluster, we establish thresholds of $\chi^2<12.8$ for our 3D spatial analysis and $\chi^2<10.6$ for our 2D kinematic analysis. For the  $\chi^2$ statistic, there is a 99.5\% probability that these two thresholds are satisfied and are plausible. If a WD meets both the 2D and 3D selection criteria, we consider it a highly reliable WD in the cluster.

\section{Binary population synthesis}
\begin{figure}[h]
    \centering
    \includegraphics[width=0.8\textwidth]{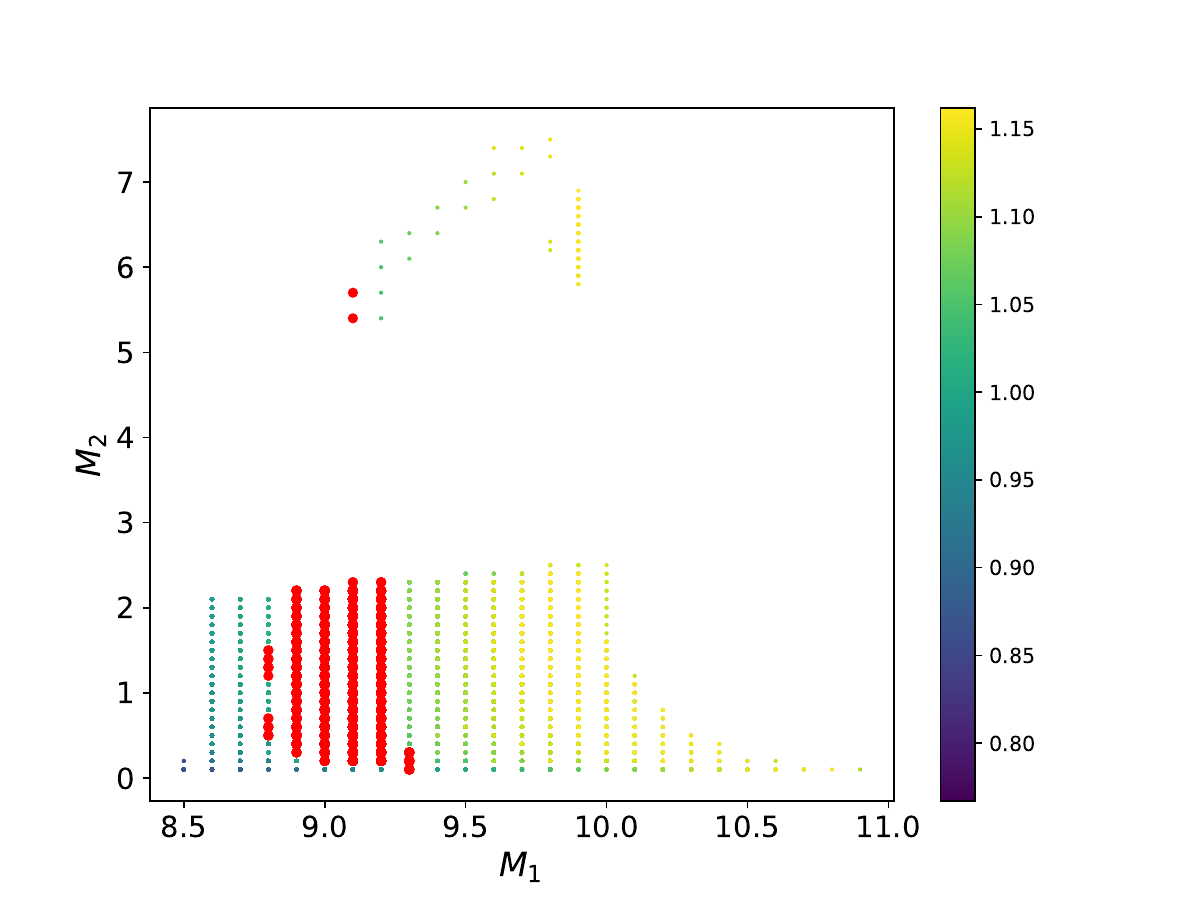}
    \caption{Initial binary mass combinations capable of producing a CO white dwarf. Red points represent those combinations that yield a CO WD with a mass between 1.02 $M_{\odot}$ and 1.06 M$_{\odot}$. The outcome of all these combinations is a single WD, with its companion having been engulfed during the most recent common envelope phase. The colors represent the final mass of the WD produced by these binary combinations. Note that each point in the figure actually represents many overlapping points, each with different initial periods, resulting in WD masses that vary accordingly (thus the color represents only the outcome for one of the periods).}
    \label{fig:s5}
\end{figure}

We utilised the rapid binary star evolution code BSE to simulate the evolutionary paths of binary systems with different parameters \citep{Hurley2002MNRAS.329..897H}. 
The primary inputs for BSE include the mass of the primary star, the mass of the secondary star, the initial period of the binary system, the orbital eccentricity, the metallicity (the elemental composition of the stars), and the integration time. For the primary star, we allowed its mass to be uniformly distributed between 5 $M_{\odot}$ and 15 $M_{\odot}$, with a step size of 0.1 $M_{\odot}$. For the secondary star, its mass ranged from 0.1 $M_{\odot}$ to 15 $M_{\odot}$, also with a step size of 0.1 $M_{\odot}$. The initial orbital periods of the binary systems varied from 1 day to 1000 days, with a step size of 1 day. The initial eccentricities included five different cases: 0, 0.25, 0.5, 0.75, and 1. Based on our fitting for RSG 5, we set the metallicity of all binary systems to solar abundance (For BSE, the solar metallicity is $Z$ = 0.02), with the integration time calculated up to 40 Myr. The above combinations of parameters resulted in approximately 76 million different initial conditions for the binary systems. Another two factors that are important for binary evolution include the efficiency of common envelope (CE) ejection, $\alpha$ and the coefficient $\eta$ for the Reimers wind mass loss. In the common envelope evolution model, $\alpha$ represents the fraction of orbital energy that is transferred to the envelope and utilised to overcome the binding energy. The default value for $\alpha$ in BSE is set to 3. This value is considered somewhat high for low-mass stars (about 0.3 \citep{Ge2024ApJ...961..202G, Ge2022ApJ...933..137G})
, but current research suggests that it can exceed 1 for massive stars \citep{Lau2022MNRAS.512.5462L} 
. In this study, we examined both cases of $\alpha=1.0$ and $3.0$. The coefficient $\eta$ for the Reimers wind mass loss quantifies the efficiency of mass loss in stars due to stellar winds \citep{Reimers1977A&A....61..217R}. 
The default value of $\eta$ is 0.5, which is close to recent observations \citep{McDonald2015MNRAS.448..502M}. 
We therefore adopt $\eta$=0.5 in our models. We then looked for cases in these binary evolution results that include a CO WD, with the companion either being absent or made up of an invisible object (such as a brown dwarf or red dwarf, neutron star, or black hole). Our results revealed no instances of a WD accompanied by a brown or red dwarf, indicating that there is no solution in which this WD can belong to a cataclysmic variable system. We identified 14 cases of WD and black hole combinations, approximately 10,000 cases of WD and neutron star pairs, as well as nearly 900,000 instances of single WDs. If we further restrict the mass of the produced WDs to between 1.02 $M_{\odot}$ and 1.06 $M_{\odot}$, in order to match our observation (1.044 $M_{\odot}$), only 20 cases of WD and neutron star combinations meet this criterion, while there are still 80,000 instances of single WDs. Thus, the combinations of WDs with compact objects are negligible compared to the instances of single WDs. In Fig. \ref{fig:s5}, we present all the initial binary mass combinations that can produce WDs within the observed mass range.

\section{Modeling binary evolution}
Using \texttt{MESA} (v24.08.1), we verified the BSE binary solution with identical initial conditions: a $9.5\,M_{\odot}$ primary, $0.2\,M_{\odot}$ secondary, solar metallicity ($Z=0.0134$), and 300-day orbital period. We terminated the simulation at 40\,Myr. For mass-transfer and CE prescriptions, we adopted the \citet{Marchant2021A&A...650A.107M} framework (their Section~2). Given the companion's extremely low mass (M-dwarf), we disabled its evolution to optimize computational efficiency. 

The evolution proceeded in two stages: (1) Starting from a pre-computed $9.5\,M_{\odot}$ ZAMS model, \texttt{MESA} naturally calculated the system's entry into a CE phase. As the CE neared complete ejection and the computation became numerically challenging, we progressively decreased the mass-loss rate to allow the simulation to continue. Finally, we terminated calculations when the envelope mass dropped below $0.001\,M_{\odot}$ to prevent numerical instabilities arising from infinitesimal timesteps. (2) We repeated the process, now using the primary star's final state from the first step, a helium star with a residual envelope, as the initial condition and adopting the binary orbital parameters calculated from the first step.

We imposed two termination criteria: (a) when the primary star reaches the WD stage, or (b) when the CE orbital period decays below 5 minutes (\texttt{MESA}'s merger threshold). The system initially met criterion (a) WD formation; we then disabled this condition to examine pre-merger binary parameters beyond the WD stage. Our final solution yields a $1.0\,M_{\odot}$ WD and $0.2\,M_{\odot}$ M-dwarf that merge during the CE phase.

\bibliography{sample7}{}
\bibliographystyle{aasjournalv7}



\end{document}